\documentclass[journal=jacsat,manuscript=article]{achemso}

\usepackage[version=3]{mhchem} 

\usepackage{booktabs} 
\usepackage{longtable}
\usepackage{array}
\usepackage[table,xcdraw]{xcolor}

\author{Dongming Li}
\affiliation[ECE]{Department of Electrical and Computer Engineering, University of Massachusetts Amherst, MA, United States.}
\author{Niamh Matthews}
\affiliation[MSE]
{Materials Science and Engineering Graduate Program, University of Massachusetts Amherst, MA, United States.}

\author{Qingchuan Sang}
\affiliation[MSE]
{Materials Science and Engineering Graduate Program, University of Massachusetts Amherst, MA, United States.}

\author{Eric Polizzi}
\email{epolizzi@engin.umass.edu}
\affiliation[ECE]
{Department of Electrical and Computer Engineering, University of Massachusetts Amherst, MA, United States.}
\alsoaffiliation[Second University]
{Department of Mathematics and Statistics, University of Massachusetts Amherst, MA, United States.}

\title[An \textsf{achemso} demo]
  {LDA-1/2 for Molecular Systems: A Real-Space Finite-Element Benchmark on the GW100 Set
  }

\abbreviations{IR,NMR,UV}

\begin{document}

\begin{abstract}
The LDA-1/2 method provides an efficient correction to semilocal density functional theory for improving ionization energies and band gaps, yet its application to molecular systems has remained limited. In this work, we present an all-electron finite-element implementation of LDA-1/2 within the \textsc{nessie} electronic-structure framework and apply it to the GW100 molecular benchmark set with systematically controllable numerical accuracy. The self-energy correction is constructed explicitly from neutral and half-ionized calculations for each molecule, avoiding the use of precomputed atomic correction potentials. The real-space finite-element formulation enables systematic convergence with respect to the discretization and provides a controlled assessment of LDA-1/2 performance. For the GW100 set, the present implementation yields a mean absolute error of $0.472$~eV and a root-mean-square error of $0.645$~eV relative to CCSD(T) reference ionization energies, substantially improving upon conventional LDA and the previously reported LAPW implementation of LDA-1/2, while achieving accuracy comparable to $G_0W_0$@PBE. Convergence tests show that third-order finite elements are sufficient to reach or approach chemical accuracy relative to higher-order calculations for the representative systems considered. The resulting corrected Hamiltonian also improves several lower-lying valence states relative to LDA, although the improvement becomes less systematic away from the HOMO. This work provides accurate LDA-1/2 benchmark data for molecular systems and establishes a rigorous
finite-element foundation for future molecular GW calculations.
\end{abstract}

\section{Introduction}
The accurate prediction of electronic excitation energies, particularly ionization potentials (IP) and electron affinities (EA), remains a central challenge in computational chemistry and condensed matter physics. While Kohn–Sham density functional theory (DFT) provides an efficient and widely used framework for computing ground-state properties, standard local and semilocal exchange–correlation approximations systematically underestimate fundamental gaps and ionization energies due to self-interaction errors and the absence of derivative discontinuities. Many-body perturbation theory within the GW approximation  and post Hartree-Fock methods, such as Coupled-Cluster, offer formally well-defined routes to quasiparticle energies, but their high computational costs motivate the development of improved low-cost alternatives and reliable reference methods.

Among such approaches, the LDA-1/2 method has emerged as a remarkably simple and effective correction to conventional DFT. Originally inspired by Slater’s transition-state and half-occupation concepts \cite{Slater1971,Slater1972-1,Slater1972-2}, the LDA-1/2 method introduces an orbital-dependent self-energy correction constructed from half-ionized atomic states, leading to significantly improved band gaps and ionization energies at a computational cost comparable to standard DFT. Over the past decade, this approach has been extensively explored and refined for periodic and extended systems, particularly semiconductors and insulators \cite{Ferreira2008,Ferreira2011,XUE2018,Xue2022}, where it has been shown to yield band gaps in close agreement with experiment and GW results while retaining the efficiency of local density or generalized gradient approximations.

Despite this success for extended systems, applications of the LDA-1/2 method to molecular systems remain relatively scarce. Only a limited number of studies have examined atoms and molecules explicitly \cite{lda12_GW100}, and these investigations have largely relied on plane-wave or localized-orbital representations. As a result, questions remain regarding the systematic accuracy, basis-set convergence, and numerical controllability of LDA-1/2 calculations for finite systems, particularly when benchmark-level precision is required.

Real-space mesh techniques, including finite-element methods (FEM), provide a powerful alternative to plane-wave and atom-centered basis-set approaches for electronic-structure calculations. By representing wave functions and potentials on real-space meshes, FEM formulations offer systematically controllable accuracy, strict variational convergence, and the flexibility to impose arbitrary boundary conditions. These properties make finite-element approaches particularly well suited for isolated molecular systems, where artificial periodicity and basis-set incompleteness can otherwise introduce subtle but non-negligible errors. However, the combination of finite-element discretizations with LDA-1/2 corrections has not yet been systematically explored.

In this work, we present a finite-element implementation of the LDA-1/2 method within the \textsc{nessie} electronic-structure framework \cite{Kestyn2020,NESSIE,LI_CPC,LI_PRB}. 
\textsc{nessie} employs full-core potentials and high-order finite elements to perform first-principles all-electron calculations. 
The FEM formulation employed in \textsc{nessie} enables systematic control of discretization errors and high-precision treatment of both wave functions and electrostatic potentials, making it a natural environment for assessing the intrinsic accuracy of LDA-1/2 corrections for molecular systems.

We apply our finite-element LDA-1/2 methodology to a broad set of isolated molecules drawn from the GW100 benchmark \cite{GW100_paper,lda12_GW100}, which provides a widely accepted reference for molecular quasiparticle energies. By carefully converging the finite-element discretization, we aim to deliver 
numerically controlled and systematically converged LDA-1/2 results,
thereby establishing a reliable reference for both the LDA-1/2 method itself and for subsequent many-body calculations. In particular, our study is motivated by recent work suggesting that LDA-1/2 orbitals and eigenvalues can serve as improved starting points for $G_0W_0$ calculations \cite{lda12gwstartingpoint}, an aspect that is especially relevant for molecular systems where starting-point dependence is often pronounced.

The goals of this paper are therefore twofold:
\begin{itemize}
    \item to assess the performance of the LDA-1/2 method for molecular systems using a systematically convergent finite-element representation;
    \item to provide high-accuracy LDA-1/2 reference data for the GW100 molecular set.
\end{itemize}

By combining the conceptual simplicity of the LDA-1/2 correction with the numerical rigor of finite-element methods, this work aims to provide the community with a robust and transferable framework for accurate molecular electronic-structure calculations.

\section{Theory and Model}

The LDA-1/2 approach can be motivated from the relation between total-energy differences and Kohn--Sham eigenvalues. The ionization potential associated with removing an electron from orbital $\alpha$ is

\begin{equation}
I_\alpha
=
E^{(\alpha)}(N-1)-E(N),
\end{equation}

where $E^{(\alpha)}(N-1)$ denotes the energy of the ionized state containing a hole in orbital $\alpha$. According to Janak's theorem \cite{Janak78},

\begin{equation}
\frac{\partial E}{\partial f_\alpha}
=
\varepsilon_\alpha(f_\alpha),
\end{equation}

with $f_\alpha$ the occupation of orbital $\alpha$. Integrating the occupation from 1 to 0 gives

\begin{equation}
E(N)-E^{(\alpha)}(N-1)
=
\int_0^1
\varepsilon_\alpha(f_\alpha)\,df_\alpha.
\end{equation}

The Slater transition-state approximation replaces this integral by its midpoint value,

\begin{equation}
E(N)-E^{(\alpha)}(N-1)
\approx
\varepsilon_\alpha
\left(
f_\alpha=\frac{1}{2}
\right)\approx -I_\alpha ,
\end{equation}

%

For the first ionization potential, $\alpha$ corresponds to the HOMO. For the exact exchange-correlation functional, the total energy is piecewise linear between integer electron numbers, such that the HOMO eigenvalue is constant over the fractional-occupation interval and satisfies

\begin{equation}
\varepsilon_{\mathrm{HOMO}}
=
E(N)-E(N-1)
=
-I.
\end{equation}

For approximate local and semilocal functionals, however, the total energy generally exhibits curvature with respect to fractional occupation. Consequently, the integer-occupation HOMO energy does not in general reproduce the ionization potential, and even the explicit half-occupation eigenvalue remains an approximation to the finite total-energy difference.

In the present implementation, the half-ionized state is therefore used to construct a correction to the effective one-electron potential rather than being used directly as the final electronic structure. The procedure consists of three self-consistent calculations.

First, a conventional neutral calculation with $N$ electrons is performed, yielding the total energy $E(N)$ and the corresponding self-consistent potential
$V_0(\mathbf{r})$.

Second, a fractional-occupation calculation is performed by removing half an electron from the HOMO. In the case of a degenerate HOMO in the neutral system, the fractional hole is distributed equally among the degenerate states. This produces the half-ionized self-consistent potential
$V_{1/2}(\mathbf{r})$.

The difference
$V_{\mathrm{S}}(\mathbf{r})
=
V_0(\mathbf{r})-V_{1/2}(\mathbf{r})$,
contains the change in the self-consistent potential induced by the creation of a half hole in the HOMO. It can therefore be interpreted as a self-energy-like correction associated with the ionization process.

A third self-consistent calculation is then performed for the neutral $N$-electron system in the presence of the correction derived from $V_{\mathrm{S}}$. The resulting one-electron problem can be written as

\begin{equation}
    \big(-\frac{1}{2}\nabla^{2}+V_{ext}(\mathbf{r})+
    V_{H}(\mathbf{r})+V_{XC}^{\mathrm{LDA}}(\mathbf{r})-V_{S}(\mathbf{r})\big)\psi_{i}(\mathbf{r})=\varepsilon_{i}^{\mathrm{LDA-1/2}}\psi_{i}(\mathbf{r})
\end{equation}







The final calculation is therefore performed at the physical integer electron number. The neutral density and orbitals are allowed to relax self-consistently in the presence of the half-ionization correction, and the resulting spectrum is not expected to be identical to that of the explicit $N-1/2$ calculation.



Since the correction potential \(V_S(\mathbf r)\) is constructed from the change in the self-consistent potential induced by half-electron removal, it contains an approximate account of the electronic response associated with the Slater transition state. When this potential is included in the final self-consistent \(N\)-electron calculation, the corrected HOMO eigenvalue is expected to provide an improved approximation to the electron-removal energy. Accordingly,

\begin{equation}
\varepsilon_{\mathrm{HOMO}}^{\mathrm{LDA-1/2}}\approx E(N)-E(N-1)=-I.
\end{equation}

%




An important advantage of the three-SCF construction is that the correction acts at the Hamiltonian level. The final calculation therefore provides not only a corrected HOMO energy but a complete set of corrected one-particle eigenvalues.
Thus, the same calculation yields the HOMO, deeper occupied states, and unoccupied states, making the approach suitable for the description of the molecular spectrum rather than only the first ionization potential.

The complete procedure is summarized schematically in Figure~\ref{fig:3scf}.

\begin{figure}[htbp]
\begin{equation}\nonumber
\boxed{
\begin{aligned}
\text{SCF 1:} \qquad
& N
\longrightarrow
V_0(\mathbf{r}),
\\[4pt]
\text{SCF 2:} \qquad
& N-\frac{1}{2}
\longrightarrow
V_{1/2}(\mathbf{r}),
\\[4pt]
\text{Correction:} \qquad
&
V_0(\mathbf{r})-V_{1/2}(\mathbf{r})
\longrightarrow
V_{S}(\mathbf{r}),
\\[4pt]
\text{SCF 3:} \qquad
& N,\; \bigl(V_{\mathrm{KS}}^{\mathrm{LDA}}-V_S\bigr)
\longrightarrow
\left\{\varepsilon_i^{\mathrm{LDA-1/2}}\right\}.
\end{aligned}
}
\end{equation}
\caption{Schematic of the three-SCF LDA-1/2 procedure. The first two SCF calculations are used to construct the self-energy correction, while the third is the corrected N-electron calculation from which the LDA-1/2 spectrum is obtained.}
\label{fig:3scf}
\end{figure}

Compared with Ref.~\citenum{lda12_GW100}, where the molecular self-energy potential is approximated as a sum of precomputed atomic contributions, we construct the correction directly from the molecule. Specifically, we obtain $V_{\mathrm{S}}(\mathbf{r})$
from neutral and half-ionized molecular calculations. This avoids the additional atomic-decomposition approximation and retains the molecule-specific response to half ionization. In both approaches, the resulting correction is then used in a self-consistent $N$-electron calculation to obtain the LDA-1/2 spectrum.

Finally, all calculations are performed within the spin-unpolarized formalism, consistent with the closed-shell character of the neutral GW100 systems. In constructing the LDA-1/2 correction, the removal of half an electron is treated in a spin-averaged manner, with equal fractional depletion of the two spin channels. The resulting \(N-\tfrac12\) calculation should therefore be regarded as a spin-averaged Slater transition-state construction used to obtain the LDA-1/2 correction potential, rather than as a spin-resolved description of the physical \(N\rightarrow N-1\) ionization process.

\section{Numerical Convergence and Parameters}

In this work, the LDA-1/2 potential is incorporated directly at the real-space level within the \textsc{nessie} finite-element framework. Because both the KS orbitals and all local potentials are represented on the same finite-element mesh, no additional approximations or basis transformations are required when adding the LDA-1/2 correction. Our \textsc{nessie} code utilizes the Finite Element Method (FEM) to solve the Kohn-Sham equations using full core potentials. The simulation volume is discretized into a mesh of elements (tetrahedra in this work). Within each element, the wavefunctions $\psi(\mathbf{r})$ are expanded using high-order Lagrange polynomials.
The primary advantage of FEM is that it combines the flexibility of a real-space grid with the variational stability of a basis-set method. Accuracy is systematically improved by decreasing the element size $h$ ($h$-refinement) or increasing the polynomial order $p$ ($p$-refinement).
In contrast to Gaussian-type orbitals, FEM provides a systematic route toward the complete-basis/discretization limit. 

To achieve systematically converged results suitable for benchmark comparisons, the following numerical settings were employed:

\begin{itemize}
    \item Mesh Discretization: A multi-scale mesh is employed, with higher element density near the nuclei to capture the rapid oscillations of the all-electron wavefunctions and a coarser mesh in the vacuum region \cite{Lehtovaara09,Kestyn2020,Levin2011}.
    \item Boundary Conditions: For isolated molecules, we impose Dirichlet boundary conditions for the wavefunction ($\psi = 0$) at a sufficiently large distance (typically $\geq$ 6 \AA) from the molecular edges to ensure that the wavefunction tails are negligible. The boundary conditions for the Hartree potential, obtained from the solution of the Poisson equation, are computed explicitly using the integral equation at the boundary surfaces. 
    \item Integration: High-order Gauss-Legendre quadrature rules are used for the assembly of the Hamiltonian and overlap matrices to maintain numerical precision.
\end{itemize}

\section{Results and discussion}



\subsection{Full GW100 Set}

The GW100 set consists of 100 molecules and atoms covering a wide range of chemical bonding types, from simple diatomics to large organic molecules and transition metal complexes. Geometries were taken from the standard GW100 database.
Table~\ref{tab:gw100} presents the HOMO ionization energies calculated with our all-electron \textsc{nessie} code using third-order finite elements (P3 FEM, cubic basis functions). Results are reported for LDA within the Perdew--Zunger parametrization, Hartree--Fock (HF), the hybrid PBE0 functional, and LDA-1/2. Additional convergence results presented further will confirm that the reported ionization energies are well converged with respect to the P3 finite-element discretization.

Experimental and CCSD(T) data are included as reference values. $G_0W_0$@PBE results computed with FHI-aims and TURBOMOLE using the def2 basis sets are included for comparison with a perturbative many-body approach \cite{GW100_paper}. LDA-1/2 results from Ref.~\citenum{lda12_GW100}, obtained using a full-potential linearized augmented-plane-wave (LAPW) method, are also included to assess the agreement between different basis representations and implementations of the same functional, in particular relative to our all-electron finite-element treatment of the full-core potential \cite{lda12_GW100}.

\begin{figure}[htbp]
    \centering
    \includegraphics[width=1\linewidth]{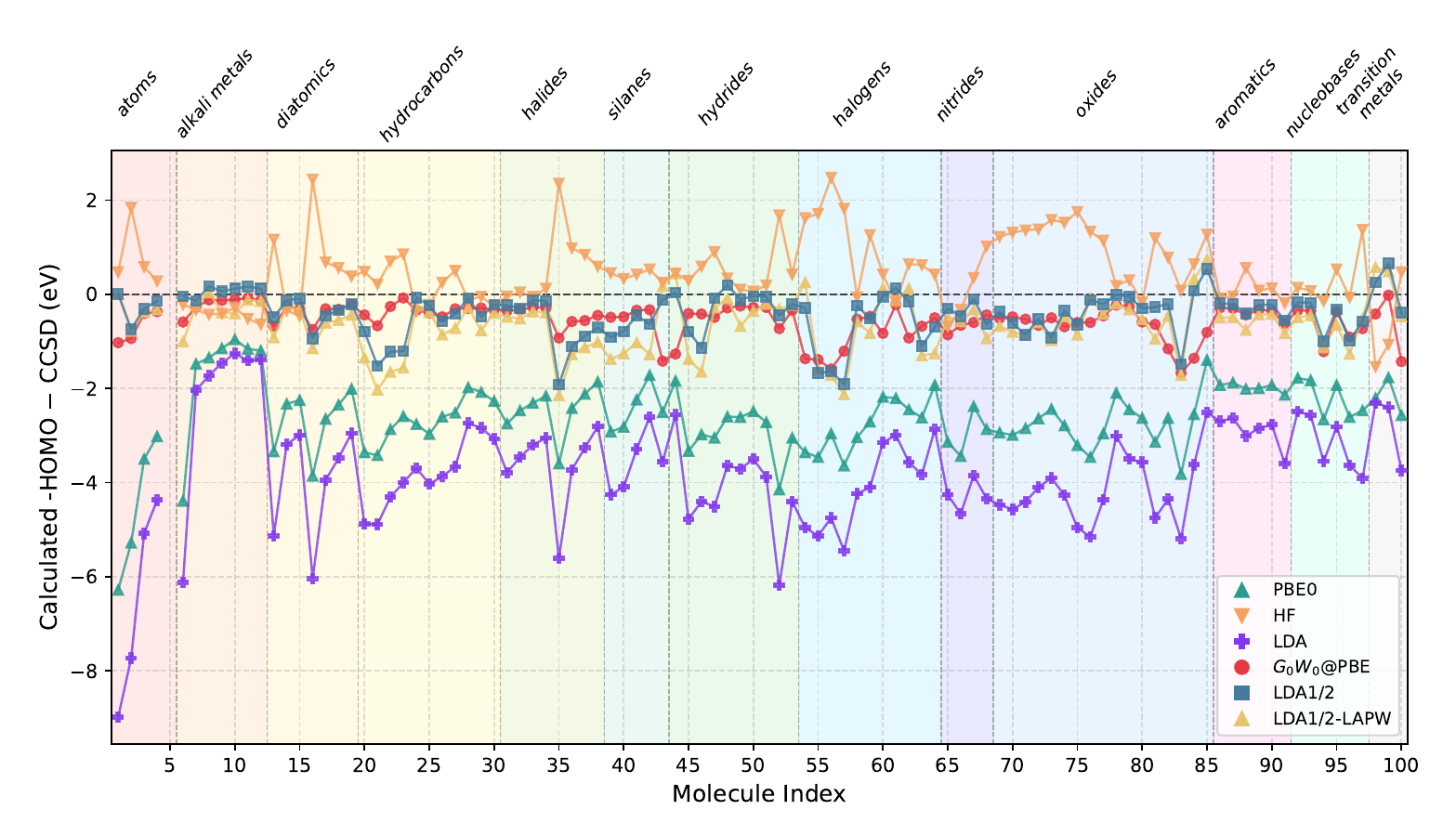}
\caption{Deviation of the HOMO ionization energies,
$-\varepsilon_{\mathrm{HOMO}}$, from the CCSD(T) reference
ionization energies for the GW100 molecular set. The molecules are ordered
according to the indexing used in Table~\ref{tab:gw100}.}
    \label{fig:gw100_1}
\end{figure}
\begin{figure}[htbp]
    \centering
    \includegraphics[width=0.65\linewidth]{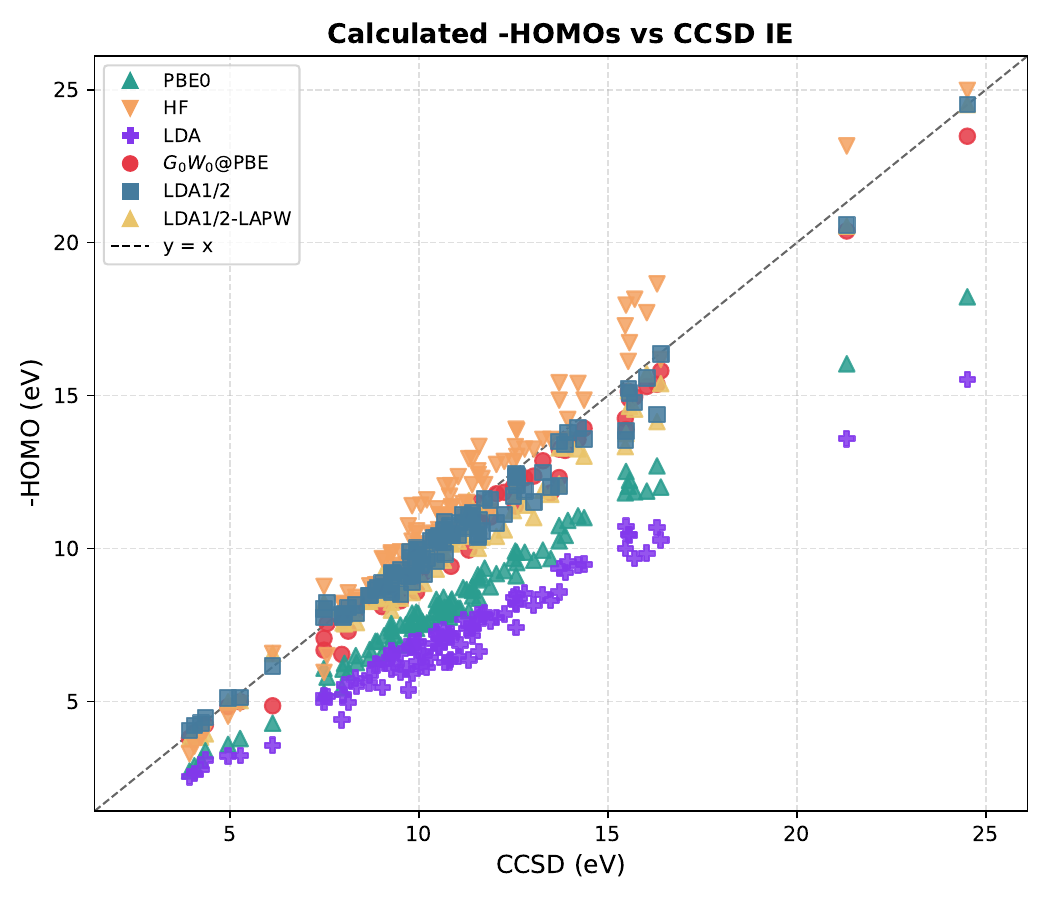}
    \caption{Calculated HOMO ionization energies,
$-\varepsilon_{\mathrm{HOMO}}$, as a function of the CCSD(T)
reference ionization energies for the GW100 molecular set. The diagonal
line indicates perfect agreement with the CCSD(T) values.}
    \label{fig:gw100_2}
\end{figure}

Figure~\ref{fig:gw100_1} presents the data from
Table~\ref{tab:gw100} as deviations of the HOMO ionization energies, $-\varepsilon_{\mathrm{HOMO}}$, from the CCSD(T) reference values, highlighting systematic trends and outliers across the GW100 set.
Figure~\ref{fig:gw100_2} shows the corresponding correlation between the calculated HOMO ionization energies and the CCSD(T) reference values.

\clearpage
\renewcommand{\arraystretch}{0.85}
\begin{small}
\begin{center}
\begin{longtable}{@{}lllccccccc@{}}
\caption{Ionization energies (eV) for the GW100 molecular benchmark set. For the single-particle methods, the reported values correspond to $-\varepsilon_{\mathrm{HOMO}}$. Experimental values are taken from Ref.~\citenum{GW100_paper}; asterisks denote vertical ionization energies.
The $G_0W_0$@PBE and CCSD(T) reference values are taken from
Refs.~\citenum{GW100_paper} and~\citenum{lda12_GW100}, respectively.
The LDA-1/2 LAPW results are taken from Ref.~\citenum{lda12_GW100}. The LDA-1/2, LDA, PBE0, and HF results (last four columns) were obtained with \textsc{nessie}.}

\label{tab:gw100}\\

\toprule
index & formula & exp    & CCSD(T)  & $G_0W_0$@PBE    & lda12-lapw & \textbf{lda12} & \textbf{pbe0}  & \textbf{hf}    & \textbf{lda}   \\ \midrule
\endfirsthead

\toprule
index & formula & exp    & CCSD(T)  & $G_0W_0$@PBE    & lda12-lapw & \textbf{lda12} & \textbf{pbe0}  & \textbf{hf}    & \textbf{lda}   \\ \midrule
\endhead

1     & He      & 24.59  & 24.51 & 23.48 & 24.51 & 24.52      & 18.23 & 24.98 & 15.53 \\
2     & Ne      & 21.56  & 21.32 & 20.38 & 20.54 & 20.58      & 16.04 & 23.16 & 13.59 \\
3     & Ar      & 15.76  & 15.54 & 15.13 & 15.16 & 15.23      & 12.04 & 16.12 & 10.46 \\
4     & Kr      & 14.00  & 13.94 & 13.57 & 13.61 & 13.80      & 10.92 & 14.22 & 9.57  \\
5     & Xe      & 12.13  &       & 12.02 & 12.05 & 12.27      & 9.70  & 12.53 & 8.56  \\
6     & H2      & 15.43  & 16.40 & 15.81 & 15.39 & 16.36      & 12.01 & 16.18 & 10.28 \\
7     & Li2     & 4.73   & 5.27  & 4.99  & 5.03  & 5.13       & 3.79  & 4.94  & 3.23  \\
8     & Na2     & 4.89   & 4.95  & 4.83  & 4.99  & 5.12       & 3.60  & 4.52  & 3.22  \\
9     & Na4     & 4.27   & 4.23  & 4.10  & 3.84  & 4.29       & 3.08  & 3.82  & 2.77  \\
10    & Na6     & 4.12   & 4.35  & 4.24  & 3.94  & 4.47       & 3.39  & 4.06  & 3.09  \\
11    & K2      & 4.06   & 4.06  & 3.98  & 3.94  & 4.22       & 2.90  & 3.54  & 2.65  \\
12    & Rb2     & 3.90   & 3.93  & 3.80  & 3.78  & 4.05       & 2.73  & 3.29  & 2.54  \\
13    & N2      & 15.58* & 15.57 & 14.89 & 14.65 & 15.08      & 12.23 & 16.73 & 10.44 \\
14    & P2      & 10.62* & 10.47 & 10.21 & 10.20 & 10.34      & 8.14  & 10.12 & 7.28  \\
15    & As2     & 10.00* & 9.78  & 9.47  & 9.46  & 9.69       & 7.53  & 9.32  & 6.79  \\
16    & F2      & 15.70* & 15.71 & 14.96 & 14.56 & 14.77      & 11.85 & 18.15 & 9.67  \\
17    & Cl2     & 11.49  & 11.41 & 11.10 & 10.80 & 10.95      & 8.76  & 12.09 & 7.46  \\
18    & Br2     & 10.51  & 10.54 & 10.22 & 9.99  & 10.20      & 8.19  & 11.10 & 7.06  \\
19    & I2      & 9.36*  & 9.51  & 9.28  & 9.07  & 9.30       & 7.50  & 9.89  & 6.56  \\
20    & CH4     & 14.35* & 14.37 & 13.93 & 13.02 & 13.58      & 11.01 & 14.85 & 9.49  \\
21    & C2H6    & 12.20* & 13.04 & 12.37 & 11.01 & 11.52      & 9.62  & 13.25 & 8.15  \\
22    & C3H8    & 11.51* & 12.05 & 11.79 & 10.40 & 10.83      & 9.18  & 12.75 & 7.74  \\
23    & C4H10   & 11.09* & 11.57 & 11.49 & 10.01 & 10.36      & 8.98  & 12.42 & 7.57  \\
24    & C2H4    & 10.68* & 10.67 & 10.33 & 10.42 & 10.60      & 7.91  & 10.30 & 6.98  \\
25    & C2H2    & 11.49* & 11.42 & 11.02 & 11.04 & 11.18      & 8.45  & 11.20 & 7.39  \\
26    & C4      & 12.54  & 11.26 & 10.78 & 10.40 & 10.68      & 8.66  & 11.51 & 7.39  \\
27    & C3H6    & 10.54* & 10.87 & 10.56 & 10.15 & 10.45      & 8.35  & 11.37 & 7.21  \\
28    & C6H6    & 9.23*  & 9.29  & 8.99  & 9.01  & 9.21       & 7.31  & 9.17  & 6.55  \\
29    & C8H8    & 8.43*  & 8.35  & 8.06  & 7.58  & 7.87       & 6.27  & 8.30  & 5.51  \\
30    & C5H6    & 8.53*  & 8.68  & 8.35  & 8.28  & 8.46       & 6.41  & 8.42  & 5.61  \\
31    & C2H3F   & 10.63* & 10.55 & 10.20 & 10.07 & 10.32      & 7.80  & 10.51 & 6.76  \\
32    & C2H3Cl  & 10.20* & 10.09 & 9.76  & 9.57  & 9.79       & 7.62  & 10.13 & 6.63  \\
33    & C2H3Br  & 9.90*  & 9.27  & 8.99  & 8.90  & 9.14       & 6.96  & 9.22  & 6.07  \\
34    & C2H3I   & 9.35*  & 9.33  & 9.04  & 8.94  & 9.19       & 7.17  & 9.45 
& 6.28  \\
35    & CF4     & 16.20* & 16.30 & 15.37 & 14.15 & 14.38      & 12.70 & 18.65 & 10.69 \\
36    & CCl4    & 11.69* & 11.56 & 10.98 & 10.28 & 10.45      & 9.14  & 12.54 & 7.83  \\
37    & CBr4    & 10.54* & 10.46 & 9.90  & 9.33  & 9.57       & 8.34  & 11.30 & 7.20  \\
38    & CI4     & 9.10*  & 9.27  & 8.82  & 8.25  & 8.56       & 7.41  &  9.86  
& 6.46  \\
39    & SiH4    & 12.82* & 12.80 & 12.31 & 11.42 & 11.89      & 9.88  & 13.25 & 8.54  \\
40    & GeH4    & 12.46* & 12.50 & 12.02 & 11.24 & 11.71      & 9.68  & 12.83 & 8.41  \\
41    & Si2H6   & 10.53* & 10.65 & 10.31 & 9.62  & 10.20      & 8.41  & 11.07 & 7.36  \\
42    & Si5H12  & 9.36*  & 9.27  & 8.94  & 7.99  & 8.63       & 7.55  & 9.80  & 6.66  \\
43    & LiH     & 7.90   & 7.96  & 6.54  & 8.13  & 7.84       & 5.45  & 8.21  & 4.40  \\
44    & KH      & 8.00   & 6.13  & 4.86  & 6.57  & 6.16       & 4.29  & 6.57  & 3.57  \\
45    & BH3     & 12.03  & 13.28 & 12.87 & 11.90 & 12.49      & 9.95  & 13.57 & 8.50  \\ 
46    & B2H6    & 11.90* & 12.26 & 11.84 & 10.61 & 11.12      & 9.28  & 12.85 & 7.86  \\
47    & NH3     & 10.82* & 10.81 & 10.32 & 10.52 & 10.73      & 7.76  & 11.71 & 6.30  \\
48    & HN3     & 10.72* & 10.68 & 10.39 & 10.59 & 10.87      & 8.08  & 11.02 & 7.04  \\
49    & PH3     & 10.59* & 10.52 & 10.27 & 9.84  & 10.41      & 7.91  & 10.62 & 6.81  \\
50    & AsH3    & 10.58* & 10.40 & 10.12 & 10.04 & 10.36      & 7.91  & 10.46 & 6.90  \\
51    & SH2     & 10.50* & 10.31 & 10.03 & 10.10 & 10.26      & 7.59  & 10.50 & 6.43  \\
52    & HF      & 16.12* & 16.03 & 15.30 & 15.71 & 15.58      & 11.88 & 17.71 & 9.85  \\
53    & HCl     & 12.79  & 12.59 & 12.25 & 12.36 & 12.38      & 9.54  & 13.01 & 8.19  \\
54    & LiF     & 11.30  & 11.32 & 9.95  & 11.57 & 11.03      & 7.96  & 12.94 & 6.36  \\
55    & MgF2    & 13.30  & 13.71 & 12.32 & 12.11 & 12.04      & 10.25 & 15.42 & 8.58  \\
56    & TiF4    &        & 15.48 & 13.89 & 13.76 & 13.84      & 12.52 & 17.96 & 10.73 \\
57    & AlF3    & 15.45* & 15.46 & 14.25 & 13.33 & 13.55      & 11.82 & 17.28 & 10.01 \\
58    & BF      & 11.00  & 11.09 & 10.56 & 10.51 & 10.85      & 8.05  & 11.03 & 6.85  \\
59    & SF4     & 12.00  & 12.59 & 12.12 & 11.76 & 12.07      & 9.88  & 13.85 & 8.49  \\
60    & KBr     & 8.82*  & 8.13  & 7.30  & 8.26  & 8.08       & 5.95  & 8.55  & 4.98  \\
61    & GaCl    & 10.07* & 9.77  & 9.55  & 9.64  & 9.90       & 7.56  & 9.60  & 6.78  \\
62    & NaCl    & 9.80*  & 9.03  & 8.10  & 9.15  & 8.88       & 6.58  & 9.67  & 5.46  \\
63    & MgCl2   & 11.80* & 11.67 & 10.99 & 10.37 & 10.57      & 9.05  & 12.29 & 7.84  \\
64    & AlI3    & 9.66*  & 9.82  & 9.32  & 8.56  & 9.12       & 7.89  & 10.24 & 6.95  \\
65    & BN      &        & 11.89 & 11.03 & 11.40 & 11.59      & 8.75  & 11.20 & 7.63  \\
66    & HCN     & 13.61* & 13.87 & 13.21 & 13.29 & 13.40      & 10.43 & 13.54 & 9.21  \\
67    & PN      & 11.88  & 11.74 & 11.14 & 11.41 & 11.64      & 9.36  & 12.09 & 7.89  \\
68    & H2NNH2  & 8.98*  & 9.72  & 9.28  & 8.78  & 9.08       & 6.85  & 10.74 & 5.38  \\
69    & H2CO    & 10.89* & 10.84 & 10.33 & 10.16 & 10.47      & 7.90  & 12.06 & 6.37  \\
70    & CH4O    & 10.96* & 11.04 & 10.56 & 10.24 & 10.42      & 8.05  & 12.35 & 6.47  \\
71    & C2H6O   & 10.64* & 10.69 & 10.16 & 9.82  & 9.83       & 7.84  & 12.05 & 6.28  \\
72    & C2H4O   & 10.24* & 10.21 & 9.55  & 9.61  & 9.68       & 7.57  & 11.59 & 6.11  \\
73    & C4H10O  & 9.61*  & 9.82  & 9.32  & 8.84  & 8.90       & 7.38  & 11.40 & 5.92  \\
74    & CH2O2   & 11.50* & 11.42 & 10.73 & 10.95 & 11.07      & 8.63  & 12.94 & 7.15  \\
75    & H2O2    & 11.70* & 11.59 & 10.99 & 10.73 & 10.94      & 8.38  & 13.34 & 6.63  \\
76    & H2O     & 12.62* & 12.57 & 11.97 & 12.48 & 12.45      & 9.11  & 13.90 & 7.42  \\
77    & CO2     & 13.77* & 13.71 & 13.25 & 13.29 & 13.50      & 10.75 & 14.85 & 9.35  \\
78    & CS2     & 10.09* & 9.98  & 9.75  & 9.78  & 9.97       & 7.88  & 10.16 & 6.97  \\
79    & OCS     & 11.19* & 11.17 & 10.91 & 10.84 & 11.12      & 8.72  & 11.47 & 7.67  \\
80    & OCSe    & 10.37* & 10.79 & 10.20 & 10.26 & 10.49      & 8.16  & 10.62 
& 7.22  \\
81    & CO      & 14.01* & 14.21 & 13.57 & 13.26 & 13.94      & 11.07 & 15.40 & 9.46  \\
82    & O3      & 12.73* & 12.55 & 11.39 & 12.09 & 12.34      & 9.92  & 13.33 & 8.20  \\
83    & SO2     & 12.50* & 13.49 & 11.82 & 11.77 & 12.01      & 9.67  & 13.57 & 8.30  \\
84    & BeO     & 10.10  & 9.94  & 8.58  & 10.26 & 10.01      & 7.39  & 10.58 & 6.32  \\
85    & MgO     & 8.76   & 7.49  & 6.68  & 8.23  & 8.03       & 6.09  & 8.76  & 4.98  \\
86    & C7H8    & 8.82*  & 8.90  & 8.61  & 8.40  & 8.71       & 6.97  & 8.82  & 6.20  \\
87    & C8H10   & 8.77*  & 8.85  & 8.55  & 8.36  & 8.64       & 6.97  & 8.81  & 6.22  \\
88    & C6F6    & 10.20* & 9.93  & 9.49  & 9.17  & 9.52       & 7.92  & 10.49 & 6.92  \\
89    & C6H5OH  & 8.75*  & 8.70  & 8.37  & 8.27  & 8.47       & 6.70  & 8.78  & 5.85  \\ 
90    & C6H5NH2 & 8.05*  & 7.99  & 7.64  & 7.57  & 7.77       & 6.06  & 8.12  & 5.22  \\
91    & C5H5N   & 9.51*  & 9.66  & 9.04  & 8.83  & 9.10       & 7.52  & 9.47  & 6.06  \\
92    & C5H5N5O & 8.24*  & 8.03  & 7.69  & 7.54  & 7.86       & 6.25  & 8.17  & 5.54  \\
93    & C5H5N5  & 8.48*  & 8.33  & 7.98  & 7.88  & 8.15       & 6.50  & 8.40  & 5.76  \\
94    & C4H5N3O & 8.94*  & 9.51  & 8.29  & 8.38  & 8.51       & 6.85  & 9.36  & 5.96  \\
95    & C5H6N2O2& 9.20*  & 9.08  & 8.71  & 8.43  & 8.76       & 7.15  & 9.61  & 6.26  \\
96    & C4H4N2O2& 9.68*  & 10.13 & 9.22  & 8.86  & 9.15       & 7.52  & 10.06 & 6.50  \\
97    & CH4N2O  & 10.15* & 10.05 & 9.32  & 9.53  & 9.47       & 7.58  & 11.42 & 6.14  \\
98    & Ag2     & 7.66   & 7.49  & 7.07  & 8.06  & 7.74       & 5.28  & 5.95  & 5.18  \\
99    & Cu2     & 7.46   & 7.57  & 7.55  & 8.05  & 8.23       & 5.80  & 6.50  & 5.17  \\
100   & CuCN    &        & 10.85 & 9.42  & 10.38 & 10.46      & 8.28  & 11.32 & 7.11  \\ \bottomrule
\end{longtable}
\end{center}
\end{small}

The overall performance of the different methods is summarized in
Table~\ref{tab:error_stats}, which reports the mean signed error (ME),
mean absolute error (MAE), root-mean-square error (RMSE), and the maximum
and minimum absolute deviations with respect to the CCSD(T) reference values.
We first note that LDA and PBE0 systematically underestimate the ionization
energies, with MEs of $-3.81$ and $-2.62$~eV, respectively. For LDA,
this behavior is associated with self-interaction error and the incorrect
asymptotic decay of the semilocal exchange-correlation potential. PBE0
partially mitigates these deficiencies through the inclusion of exact
exchange, but retains residual self-interaction and an incorrect asymptotic
potential, leading to HOMO eigenvalues that remain too high in energy.
In contrast, HF systematically overestimates the ionization energies, reflecting the neglect of electron correlation.
Despite this tendency toward overestimation, HF shows performance competitive with $G_0W_0$, which instead tends to underestimate the ionization energies.

Table~\ref{tab:error_stats} shows that
the present LDA-1/2 implementation
substantially reduces the error relative to conventional LDA and PBE0, with an
MAE of  $0.472$~eV and an RMSE of  $0.645$~eV.
Its performance is also improved relative to the LAPW LDA-1/2 results and is
comparable to $G_0W_0$@PBE for this benchmark set.
While the \textsc{nessie} finite-element and LAPW implementations of LDA-1/2 show similar overall trends across the molecular set, the present finite-element results exhibit noticeably smaller errors. Although the two approaches employ different basis representations, both are assumed to be sufficiently converged. 
Two important methodological differences that may contribute to this improvement are
in the treatment of the core region and in the construction of the self-energy correction potential. In particular, our all-electron full-core finite-element formulation avoids the LAPW partitioning of space, while the self-energy potential is generated explicitly for each molecule rather than assembled from precomputed atomic
contributions.

\begin{table}[htbp]
\caption{Error statistics for the values reported in Table~\ref{tab:gw100}, where ME is the mean signed error, MAE the mean absolute error, RMSE the root-mean-square error, and MAX and MIN the maximum and minimum absolute deviations, respectively. The qsGW statistics from Ref.~\citenum{caruso2016} are also included for comparison.}
\label{tab:error_stats}
\begin{tabular}{@{}llllll@{}}
\toprule
functional  & ME      & MAE     & RMSE    & MAX  & MIN  \\ \midrule
qsGW\cite{caruso2016}        & 0.1462  & 0.2181 & 0.2916 & 0.95 & 0    \\
\ce{G0W0}@PBE\cite{GW100_paper} & -0.5524 & 0.5524 & 0.6584  & 1.67 & 0.02 \\
lda12       & -0.4215 & 0.4722 & 0.6452 & 1.92 & 0.01 \\
lda12-lapw\cite{lda12_GW100}  & -0.6346 & 0.7004  & 0.8565 & 2.15 & 0    \\
hf          & 0.4718  & 0.6601  & 0.8811 & 2.48 & 0.04 \\
pbe0        & -2.624 & 2.624 & 2.744 & 6.28 & 0.96 \\
lda         & -3.812 & 3.812 & 4.001  & 8.98 & 1.26 \\ \bottomrule
\end{tabular}
\end{table}

The LDA-1/2 method is obtained from a small number of self-consistent calculations and remains computationally inexpensive compared with many-body perturbation theory approaches.
Although LDA-1/2 achieves an MAE comparable to $G_0W_0$@PBE, its accuracy remains below that of more advanced self-consistent GW schemes such as qsGW. In $G_0W_0$, the quasiparticle correction is evaluated perturbatively from a fixed DFT starting point, and the resulting energies can therefore retain a significant starting-point dependence. By contrast, qsGW iteratively updates the effective one-particle description until self-consistency is reached, which substantially reduces this dependence and generally improves the quasiparticle energies.


\subsubsection{Convergence and basis set accuracy}

\begin{figure}
    \centering
    \includegraphics[width=1\linewidth]{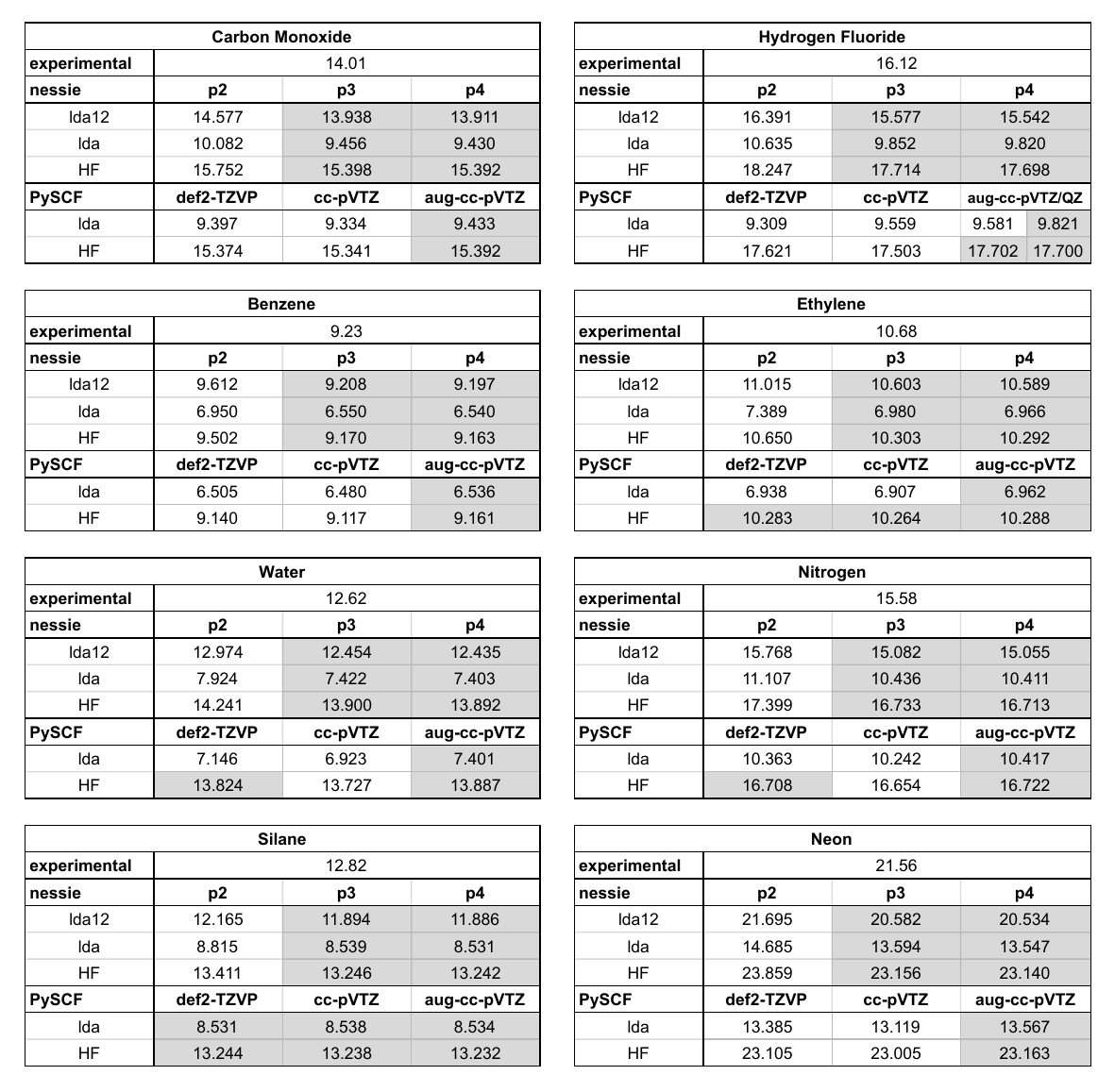}
\caption{Convergence of the HOMO ionization energies for a representative set of molecules and atoms (CO, HF, C6H6, C2H4, H2O, N2, SiH4, Ne) calculated with \textsc{nessie} using P2, P3, and P4 finite-element basis orders, together with PySCF\cite{pyscf18} results obtained using def2-TZVP, cc-pVTZ, and aug-cc-pVTZ(/pVQZ) Gaussian basis sets. The shaded gray region indicates deviations within or close to chemical accuracy ($\pm 0.043$~eV), measured relative to the most converged result available for each basis-set family.}
    \label{fig:PFEM}
\end{figure}

Figure~\ref{fig:PFEM} illustrates the convergence of the HOMO ionization
energies for a representative set of eight molecular systems calculated with
\textsc{nessie} using polynomial finite-element basis orders P2, P3, and P4,
where PN denotes a finite-element basis of polynomial order $N$. The P3 results for LDA, LDA-1/2, and HF are already converged with respect to P4 to within or close to chemical accuracy ($\sim 0.043$~eV) across all molecules considered.
The additional computational cost of P4 therefore provides only a negligible
improvement in the calculated ionization energies. These convergence results
motivate the use of P3 finite elements for the GW100 benchmark calculations
reported in Table~\ref{tab:gw100}.

Results from PySCF~\cite{pyscf18} using three Gaussian-type basis sets are also included for comparison: def2-TZVP, a general-purpose triple-$\zeta$ basis with polarization functions~\cite{weigend2005}; cc-pVTZ, a correlation-consistent triple-$\zeta$ basis designed for systematic convergence toward the complete-basis-set limit~\cite{kendall1992}; and aug-cc-pVTZ, which augments cc-pVTZ with diffuse functions to improve the description of the long-range electron density~\cite{kendall1992}. 
These reference values agree closely with the converged \textsc{nessie} P3/P4 results, with hydrogen fluoride being the notable exception, for which a larger aug-cc-pVQZ basis is required to reach comparable convergence. This larger basis-set requirement is consistent with the strongly polar character of hydrogen fluoride and the need for a more flexible description of its diffuse electron density.
For several molecules, the cc-pVTZ results exhibit larger residual deviations, indicating that this basis is not uniformly converged across the set. The comparison therefore supports the interpretation that the remaining discrepancies primarily arise from basis-set incompleteness in the Gaussian calculations, while further demonstrating the systematic convergence of the finite-element basis.

\subsection{Deviations from experimental spectrum}

\begin{figure}
    \centering
    \includegraphics[width=.48\linewidth]{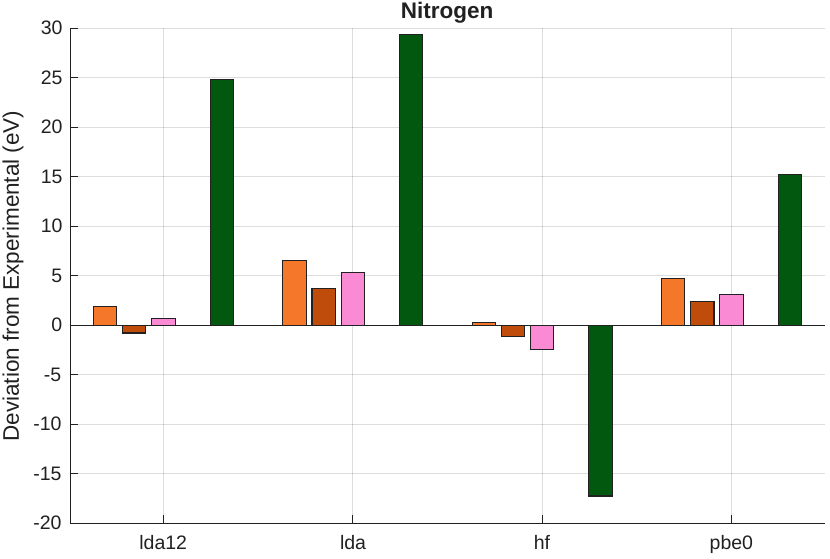} \hfill
    \includegraphics[width=.48\linewidth]{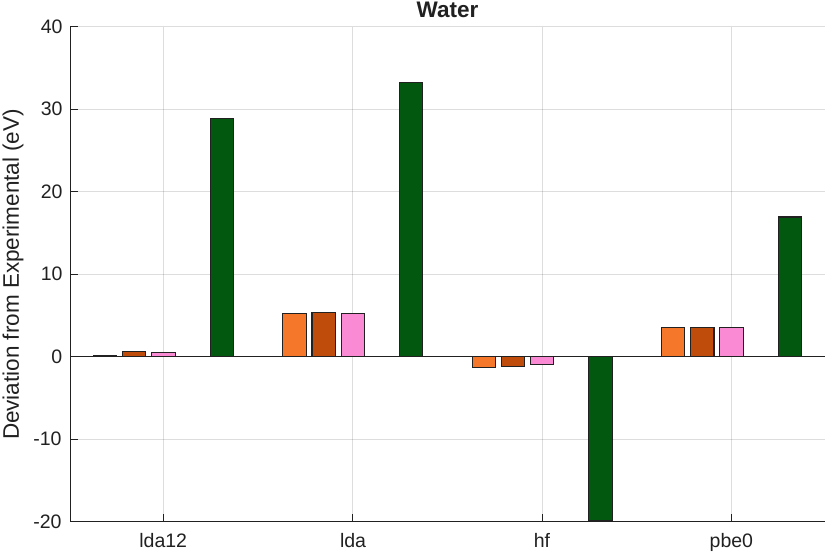} 
    \\[\smallskipamount]
    \includegraphics[width= .48\linewidth]{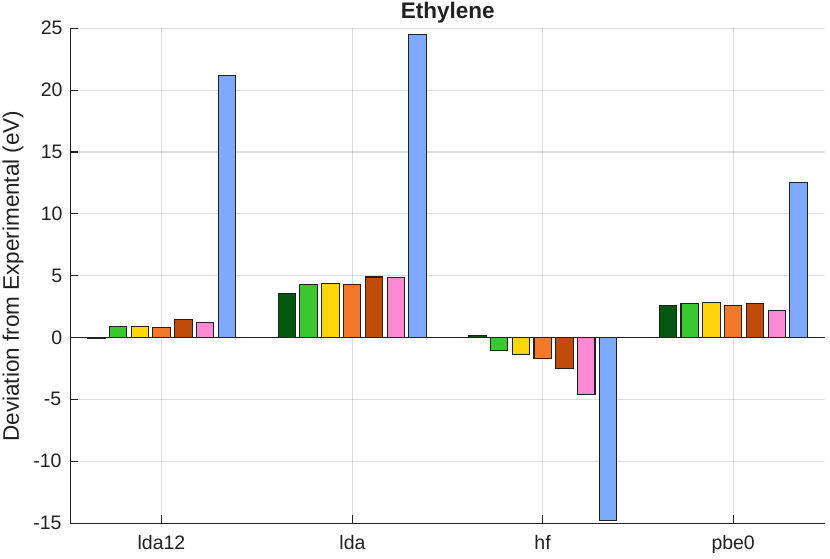} \hfill
    \includegraphics[width=.48\linewidth]{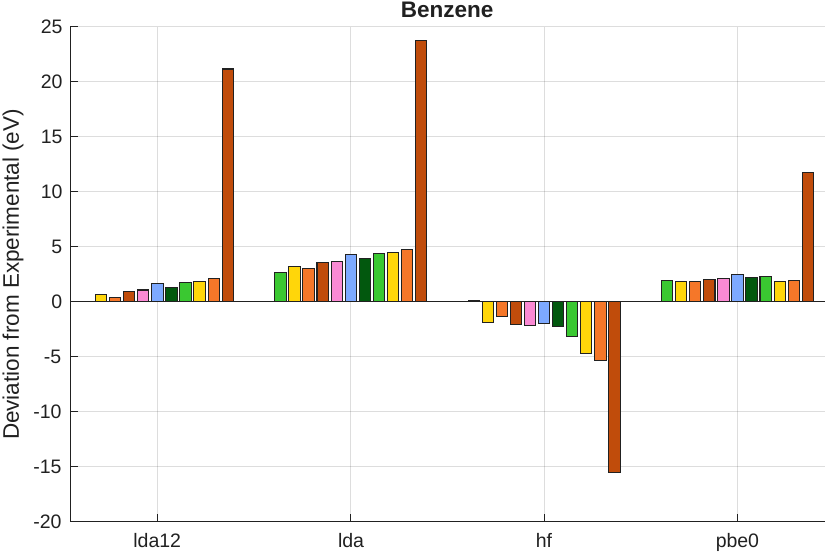} 
    \\[\smallskipamount]
    \includegraphics[width=.48\linewidth]{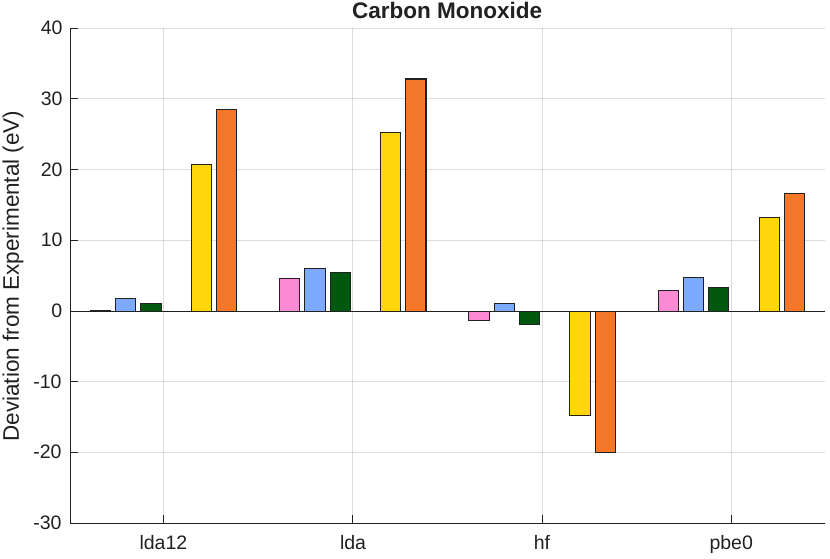} \hfill
    \includegraphics[width=.48\linewidth]{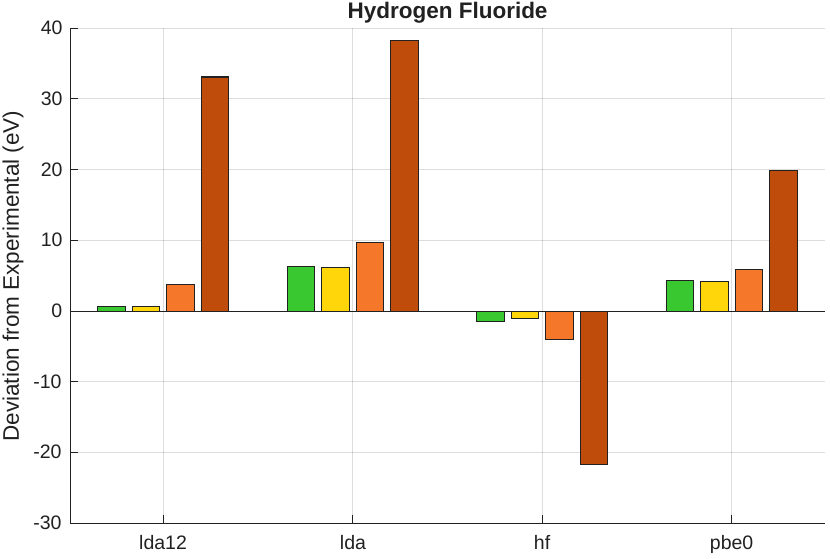}
    \caption{Lower energy levels from (L-R) HOMO, HOMO-1, HOMO-2, etc for $\rm N_2$, $\rm H_2O$, $\rm C_2H_4$, $\rm C_6H_6$, $\rm CO$, and $\rm HF$ compared against experimental data from Ref. \citenum{chong2002}} (valence states) and Ref. \citenum{jolly1984} (core states).
    \label{fig:spectrum}
\end{figure}

Figure~\ref{fig:spectrum} shows the deviation from experimental values of lower-lying valence states  for six selected molecules computed using \textsc{nessie}. The carbon monoxide $3\sigma$ orbital, water $2a1$ orbital, and nitrogen $2\sigma u$ orbital were excluded due to the lack of experimental data  \cite{marieloos2024}.
Low-lying valence states are more difficult to characterize accurately, as electron removal from these states can involve significant many-body effects and give rise to satellite structures in the photoemission spectrum~\cite{cederbaum1986}.

LDA-1/2 shows improvement over LDA for these lower energy levels, though the improvement is less pronounced than that observed for the HOMO. This is expected, as LDA-1/2's self-energy correction 
modifies the effective Kohn–Sham potential and therefore affects all orbitals,
but the correction is parameterized for the HOMO and is not optimized for deeper eigenvalues.
Consequently, the improvement becomes less systematic for deeper states than for the HOMO.


Table~\ref{tab:pyscfdata} summarizes the results shown in Figure~\ref{fig:spectrum} by reporting the mean signed error (ME) for the HOMO and several lower-lying valence states obtained with LDA-1/2, LDA, HF, and PBE0. The results show that LDA-1/2 provides a substantial reduction in the error not only for the HOMO, but also for the first few states below it.

To further assess the accuracy of the finite-element basis, Table~\ref{tab:pyscfdata} also includes PySCF results obtained with the aug-cc-pVTZ basis for LDA and HF. These values are in very good agreement with the corresponding \textsc{nessie} results for the same set of molecules, providing an additional validation of the basis-set convergence.

\begin{table}[htbp]
\caption{Mean signed errors (MEs, in eV) for the HOMO and selected lower-lying valence states obtained with LDA-1/2, PBE0, LDA, and HF using the \textsc{nessie} finite-element implementation. For comparison, PySCF results obtained with the aug-cc-pVTZ basis are also reported for LDA and HF.}
\label{tab:pyscfdata}
\begin{tabular}{@{}lcc|cc|cc@{}}
  & fem   & fem   & fem    & aug-cc-pVTZ & fem    & aug-cc-pVTZ \\
  & lda12 & pbe0  & lda    & lda      & hf     & hf          \\
\midrule
homo-0 & 0.447  & 3.340  & 4.808  & 4.829             & -0.613  & -0.600          \\
homo-1 & 0.651  & 3.241  & 4.778  & 4.798             & -0.874  & -0.955          \\
homo-2 & 1.226  & 3.403  & 5.501  & 5.528             & -2.027  & -2.008         
\end{tabular}
\end{table}





\section{Conclusion}

In this work, we have presented an all-electron, full-core finite-element
implementation of the LDA-1/2 method for molecular systems within
\textsc{nessie}. In contrast to implementations based on precomputed atomic
self-energy potentials, the present approach constructs the LDA-1/2 correction
directly from neutral and half-ionized calculations of each molecule. The
resulting correction is then incorporated into a final self-consistent
$N$-electron calculation, providing both an improved HOMO ionization energy
and a corrected molecular spectrum.

The method was benchmarked on the GW100 set using systematically converged
finite-element calculations. Relative to CCSD(T) reference ionization
energies, the present LDA-1/2 implementation yields an MAE of $0.472$~eV
and an RMSE of $0.645$~eV, representing a substantial improvement over
conventional LDA and PBE0. The accuracy is comparable to $G_0W_0$@PBE
and improves upon the previously reported LAPW implementation of LDA-1/2,
which yields an MAE of approximately $0.70$~eV for the same benchmark.
These results demonstrate that the molecule-specific construction of the
self-energy potential, combined with an all-electron real-space treatment,
provides an accurate realization of LDA-1/2 for finite systems.

The finite-element discretization was found to converge systematically, with
P3 results already within or close to chemical accuracy relative to P4 for
the representative systems considered. Independent comparisons with
Gaussian-basis PySCF calculations further support the numerical convergence
of the finite-element results. Beyond the HOMO, LDA-1/2 also improves the
first few lower-lying valence states relative to LDA, although the improvement
becomes less systematic for deeper levels, as expected from a correction
constructed primarily to reproduce the HOMO electron-removal energy.

Although the present work focuses primarily on accuracy, the computational
cost of the three-SCF LDA-1/2 procedure should also be considered. In
\textsc{nessie}, this overhead is mitigated by the highly parallel FEAST
eigensolver~\cite{polizzi09,kestyn16}. 
Interestingly, FEAST can be formulated to directly address the SCF Kohn--Sham problem by recasting it as a nonlinear eigenvector problem, as first introduced in Ref.~\citenum{gavin13}.
All results presented here were obtained using recent advances in the nonlinear FEAST formulation within \textsc{nessie}, which eliminate conventional outer SCF iterations~\cite{unpublished}. As a result, the three-SCF structure of the present LDA-1/2 method does not translate into a threefold computational overhead relative to a standard SCF calculation.
Together with the systematically convergent finite-element representation,
these developments make the present framework particularly attractive for
large-scale molecular calculations. Extending \textsc{nessie} to GW-level
quasiparticle corrections~\cite{LI_PRB} using the obtained LDA-1/2 spectrum represents a natural direction for future work.

\bibliography{LMP2026_JCTC}

@article{lda12_GW100,
author = {Rodrigues
Pela, Ronaldo and Gulans, Andris and Draxl, Claudia},
title = {The LDA-1/2 Method Applied to Atoms and Molecules},
journal = {Journal of Chemical Theory and Computation},
volume = {14},
number = {9},
pages = {4678-4686},
year = {2018},
doi = {10.1021/acs.jctc.8b00518},
    note ={PMID: 30119607},
URL = { 
        https://doi.org/10.1021/acs.jctc.8b00518
},
eprint = { 
        https://doi.org/10.1021/acs.jctc.8b00518
}
}

@article{Ferreira2008,
  title = {Approximation to density functional theory for the calculation of band gaps of semiconductors},
  author = {Ferreira, Luiz G. and Marques, Marcelo and Teles, Lara K.},
  journal = {Phys. Rev. B},
  volume = {78},
  issue = {12},
  pages = {125116},
  numpages = {9},
  year = {2008},
  month = {Sep},
  publisher = {American Physical Society},
  doi = {10.1103/PhysRevB.78.125116},
  url = {https://link.aps.org/doi/10.1103/PhysRevB.78.125116}
}

@article{Ferreira2011,
    author = {Ferreira, Luiz G. and Marques, Marcelo and Teles, Lara K.},
    title = {Slater half-occupation technique revisited: the LDA-1/2 and GGA-1/2 approaches for atomic ionization energies and band gaps in semiconductors},
    journal = {AIP Advances},
    volume = {1},
    number = {3},
    pages = {032119},
    year = {2011},
    month = {07},
    issn = {2158-3226},
    doi = {10.1063/1.3624562},
    url = {https://doi.org/10.1063/1.3624562},
    eprint = {https://pubs.aip.org/aip/adv/article-pdf/doi/10.1063/1.3624562/12995404/032119_1_online.pdf},
}

@article{XUE2018,
title = {Improved LDA-1/2 method for band structure calculations in covalent semiconductors},
journal = {Computational Materials Science},
volume = {153},
pages = {493-505},
year = {2018},
issn = {0927-0256},
doi = {https://doi.org/10.1016/j.commatsci.2018.06.036},
url = {https://www.sciencedirect.com/science/article/pii/S0927025618304166},
author = {Kan-Hao Xue and Jun-Hui Yuan and Leonardo R.C. Fonseca and Xiang-Shui Miao}
}

@article{Xue2022,
doi = {10.1088/1361-648X/ac829d},
url = {https://doi.org/10.1088/1361-648X/ac829d},
year = {2022},
month = {aug},
publisher = {IOP Publishing},
volume = {34},
number = {40},
pages = {403001},
author = {Mao, Ge-Qi and Yan, Zhao-Yi and Xue, Kan-Hao and Ai, Zhengwei and Yang, Shengxin and Cui, Hanli and Yuan, Jun-Hui and Ren, Tian-Ling and Miao, Xiangshui},
title = {DFT-1/2 and shell DFT-1/2 methods: electronic structure calculation for semiconductors at LDA complexity},
journal = {Journal of Physics: Condensed Matter}
}

@ARTICLE{Kestyn2020,
  author={Kestyn, James and Polizzi, Eric},
  journal={IEEE Nanotechnology Magazine}, 
  title={From Fundamental First-Principle Calculations to Nanoengineering Applications: A Review of the NESSIE Project}, 
  year={2020},
  volume={14},
  number={6},
  pages={52-C3},
  doi={10.1109/MNANO.2020.3024387}}

@article{LI_CPC,
title = {A method of calculating bandstructure in real-space with application to all-electron and full potential},
journal = {Computer Physics Communications},
volume = {295},
pages = {109014},
year = {2024},
issn = {0010-4655},
doi = {https://doi.org/10.1016/j.cpc.2023.109014},
url = {https://www.sciencedirect.com/science/article/pii/S0010465523003594},
author = {Dongming Li and James Kestyn and Eric Polizzi}
}

@article{LI_PRB,
  title = {Nonlinear eigenvalue algorithm for $GW$ quasiparticle equations},
  author = {Li, Dongming and Polizzi, Eric},
  journal = {Phys. Rev. B},
  volume = {111},
  issue = {4},
  pages = {045137},
  numpages = {8},
  year = {2025},
  month = {Jan},
  publisher = {American Physical Society},
  doi = {10.1103/PhysRevB.111.045137},
  url = {https://link.aps.org/doi/10.1103/PhysRevB.111.045137}
}

@article{GW100_paper,
author = {van Setten, Michiel J. and Caruso, Fabio and Sharifzadeh, Sahar and Ren, Xinguo and Scheffler, Matthias and Liu, Fang and Lischner, Johannes and Lin, Lin and Deslippe, Jack R. and Louie, Steven G. and Yang, Chao and Weigend, Florian and Neaton, Jeffrey B. and Evers, Ferdinand and Rinke, Patrick},
title = {GW100: Benchmarking G0W0 for Molecular Systems},
journal = {Journal of Chemical Theory and Computation},
volume = {11},
number = {12},
pages = {5665-5687},
year = {2015},
doi = {10.1021/acs.jctc.5b00453},
    note ={PMID: 26642984},
URL = { 
        https://doi.org/10.1021/acs.jctc.5b00453
},
eprint = { 
        https://doi.org/10.1021/acs.jctc.5b00453
}
}

@article{Slater1971,
author = {Slater, J. C.},
title = {Treatment of exchange in atomic, molecular, and solid-state theory},
journal = {International Journal of Quantum Chemistry},
volume = {5},
number = {S5},
pages = {403-409},
doi = {https://doi.org/10.1002/qua.560050848},
url = {https://onlinelibrary.wiley.com/doi/abs/10.1002/qua.560050848},
eprint = {https://onlinelibrary.wiley.com/doi/pdf/10.1002/qua.560050848},
year = {1971}
}

@article{Slater1972-1,
  title = {Self-Consistent-Field $X\ensuremath{\alpha}$ Cluster Method for Polyatomic Molecules and Solids},
  author = {Slater, J. C. and Johnson, K. H.},
  journal = {Phys. Rev. B},
  volume = {5},
  issue = {3},
  pages = {844--853},
  numpages = {0},
  year = {1972},
  month = {Feb},
  publisher = {American Physical Society},
  doi = {10.1103/PhysRevB.5.844},
  url = {https://link.aps.org/doi/10.1103/PhysRevB.5.844}
}

@incollection{Slater1972-2,
title = {Statistical Exchange-Correlation in the Self-Consistent Field},
editor = {Per-Olov Löwdin},
series = {Advances in Quantum Chemistry},
publisher = {Academic Press},
volume = {6},
pages = {1-92},
year = {1972},
issn = {0065-3276},
doi = {https://doi.org/10.1016/S0065-3276(08)60541-9},
url = {https://www.sciencedirect.com/science/article/pii/S0065327608605419},
author = {John C. Slater}
}

@article{lda12gwstartingpoint,
  title = {Probing the LDA-1/2 method as a starting point for ${G}_{0}{W}_{0}$ calculations},
  author = {Rodrigues Pela, Ronaldo and Werner, Ute and Nabok, Dmitrii and Draxl, Claudia},
  journal = {Phys. Rev. B},
  volume = {94},
  issue = {23},
  pages = {235141},
  numpages = {9},
  year = {2016},
  month = {Dec},
  publisher = {American Physical Society},
  doi = {10.1103/PhysRevB.94.235141},
  url = {https://link.aps.org/doi/10.1103/PhysRevB.94.235141}
}

@article{caruso2016,
author = {Caruso, Fabio and Dauth, Matthias and Setten, Michiel J. and Rinke, Patrick},
title = {Benchmark of GW Approaches for the GW100 Test Set},
journal = {Journal of Chemical Theory and Computation},
year = {2016},
doi = {10.1021/acs.jctc.6b00774},
url = {http://dx.doi.org/10.1021/acs.jctc.6b00774}
}

@article{kendall1992,
    author = {Kendall, Rick A. and Dunning, Thom H., Jr. and Harrison, Robert J.},
    title = {Electron affinities of the first‐row atoms revisited. Systematic basis sets and wave functions},
    journal = {The Journal of Chemical Physics},
    volume = {96},
    number = {9},
    pages = {6796-6806},
    year = {1992},
    month = {05},
    issn = {0021-9606},
    doi = {10.1063/1.462569},
    url = {https://doi.org/10.1063/1.462569},
    eprint = {https://pubs.aip.org/aip/jcp/article-pdf/96/9/6796/18998924/6796_1_online.pdf},
}

@Article{weigend2005,
author ="Weigend, Florian and Ahlrichs, Reinhart",
title  ="Balanced basis sets of split valence{,} triple zeta valence and quadruple zeta valence quality for H to Rn: Design and assessment of accuracy",
journal  ="Phys. Chem. Chem. Phys.",
year  ="2005",
volume  ="7",
issue  ="18",
pages  ="3297-3305",
publisher  ="The Royal Society of Chemistry",
doi  ="10.1039/B508541A",
url  ="http://dx.doi.org/10.1039/B508541A"}

@article{jolly1984,
title = {Core-electron binding energies for gaseous atoms and molecules},
journal = {Atomic Data and Nuclear Data Tables},
volume = {31},
number = {3},
pages = {433-493},
year = {1984},
issn = {0092-640X},
doi = {https://doi.org/10.1016/0092-640X(84)90011-1},
url = {https://www.sciencedirect.com/science/article/pii/0092640X84900111},
author = {W.L. Jolly and K.D. Bomben and C.J. Eyermann}
}

@article{chong2002,
    author = {Chong, D. P. and Gritsenko, O. V. and Baerends, E. J.},
    title = {Interpretation of the Kohn–Sham orbital energies as approximate vertical ionization potentials},
    journal = {The Journal of Chemical Physics},
    volume = {116},
    number = {5},
    pages = {1760-1772},
    year = {2002},
    month = {02},
    issn = {0021-9606},
    doi = {10.1063/1.1430255},
    url = {https://doi.org/10.1063/1.1430255},
    eprint = {https://pubs.aip.org/aip/jcp/article-pdf/116/5/1760/19314820/1760_1_online.pdf},
}

@article{marieloos2024,
author = {Marie, Antoine and Loos, Pierre-Fran{\c{c}}ois},
title = {Reference Energies for Valence Ionizations and Satellite Transitions},
journal = {Journal of Chemical Theory and Computation},
volume = {20},
number = {11},
pages = {4751-4777},
year = {2024},
doi = {10.1021/acs.jctc.4c00216},
    note ={PMID: 38776293},
url = {https://doi.org/10.1021/acs.jctc.4c00216},
eprint = {https://doi.org/10.1021/acs.jctc.4c00216}

}

@article{cederbaum1986,
  title={Correlation effects in the ionization of molecules: breakdown of the molecular orbital picture},
  author={Cederbaum, LS and Domcke, W and Schirmer, J and Niessen, W von},
  journal={Advances in chemical physics},
  pages={115--159},
  year={1986},
  publisher={Wiley Online Library}
}

@misc{NESSIE,
author ={NESSIE},
title = {High-Performance first-principle simulations},
howpublished = "\url{http://www.nessie-code.org/}",
year = {2022-present}
}

@article{Janak78,
  title = {Proof that $\frac{\ensuremath{\partial}E}{\ensuremath{\partial}{n}_{i}}=\ensuremath{\epsilon}$ in density-functional theory},
  author = {Janak, J. F.},
  journal = {Phys. Rev. B},
  volume = {18},
  issue = {12},
  pages = {7165--7168},
  numpages = {0},
  year = {1978},
  month = {Dec},
  publisher = {American Physical Society},
  doi = {10.1103/PhysRevB.18.7165},
  url = {https://link.aps.org/doi/10.1103/PhysRevB.18.7165}
}

@article{Levin2011,
  title={FEAST fundamental framework for electronic structure calculations: Reformulation and solution of the muffin-tin problem},
  author={Alan R. Levin and Deyin Zhang and Eric Polizzi},
  journal={Comput. Phys. Commun.},
  year={2011},
  volume={183},
  pages={2370-2375},
  url={https://api.semanticscholar.org/CorpusID:34658566}
}

@article{pyscf18,
  author  = {Sun, Qiming and Berkelbach, Timothy C. and Blunt, Nick S. and Booth, George H. and Guo, Sheng and Li, Zhendong and Liu, Junzi and McClain, James D. and Sayfutyarova, Elvira R. and Sharma, Sandeep and Wouters, Sebastian and Chan, Garnet Kin-Lic},
  title   = {PySCF: the Python-based simulations of chemistry framework},
  journal = {WIREs Computational Molecular Science},
  volume  = {8},
  number  = {1},
  pages   = {e1340},
  year    = {2018},
  doi     = {10.1002/wcms.1340}
}

@article{Lehtovaara09,
    author = {Lehtovaara, Lauri and Havu, Ville and Puska, Martti},
    title = {All-electron density functional theory and time-dependent density functional theory with high-order finite elements},
    journal = {The Journal of Chemical Physics},
    volume = {131},
    number = {5},
    pages = {054103},
    year = {2009},
    month = {08},
    issn = {0021-9606},
    doi = {10.1063/1.3176508},
    url = {https://doi.org/10.1063/1.3176508},
    eprint = {https://pubs.aip.org/aip/jcp/article-pdf/doi/10.1063/1.3176508/15651531/054103_1_online.pdf},
}

@article{gavin13,
    author = {Gavin, B. and Polizzi, E.},
    title = {Non-linear eigensolver-based alternative to traditional SCF methods},
    journal = {The Journal of Chemical Physics},
    volume = {138},
    number = {19},
    pages = {194101},
    year = {2013},
    month = {05},
    issn = {0021-9606},
    doi = {10.1063/1.4804419},
    url = {https://doi.org/10.1063/1.4804419},
    eprint = {https://pubs.aip.org/aip/jcp/article-pdf/doi/10.1063/1.4804419/13865253/194101_1_online.pdf},
}

@misc{unpublished,
author = {Sang, Qingchuan and Matthews, Niamh and Li, Dongming and Eric Polizzi},
  author       = {Author, A. and Author, B.},
  title        = {Nonlinear FEAST for Self-Consistent-Field Electronic-Structure Calculations},
  howpublished = {Manuscript in preparation},
  year         = {2026}
}

@article{Polizzi09,
  title = {Density-matrix-based algorithm for solving eigenvalue problems},
  author = {Polizzi, Eric},
  journal = {Phys. Rev. B},
  volume = {79},
  issue = {11},
  pages = {115112},
  numpages = {6},
  year = {2009},
  month = {Mar},
  publisher = {American Physical Society},
  doi = {10.1103/PhysRevB.79.115112},
  url = {https://link.aps.org/doi/10.1103/PhysRevB.79.115112}
}

@INPROCEEDINGS{Kestyn16,
  author={Kestyn, James and Kalantzis, Vasileios and Polizzi, Eric and Saad, Yousef},
  booktitle={SC '16: Proceedings of the International Conference for High Performance Computing, Networking, Storage and Analysis}, 
  title={PFEAST: A High Performance Sparse Eigenvalue Solver Using Distributed-Memory Linear Solvers}, 
  year={2016},
  volume={},
  number={},
  pages={178-189},
  doi={10.1109/SC.2016.15}}

\end{document}